\documentclass[preprint]{article}
\usepackage{graphicx}

\newcommand{\be}{\begin{equation}}
\newcommand{\ee}{\end{equation}}
\newcommand{\ba}{\begin{eqnarray}}
\newcommand{\ea}{\end{eqnarray}}

\begin{document}

\setlength{\textheight}{8.5 in}
\begin{titlepage}

\hfill January 23, 2018\\
\vskip 5mm

\vskip 1mm
\vskip 20mm

\begin{center}
{\Large\bf {Conformal Bootstrap Analysis for Yang-Lee Edge Singularity  }}
\vskip 6mm
 S. Hikami 

\vskip 5mm

Mathematical and Theoretical Physics Unit,
Okinawa Institute of Science and Technology Graduate University,
Okinawa, Onna 904-0495, Japan; hikami@oist.jp

\end{center}
\vskip 20mm

{\bf Abstract}
\vskip 3mm
The Yang-Lee edge singularity is investigated by the determinant method of the conformal field theory.   The  $3\times 3$ minors are used for the evaluation of the scale dimension.  The agreement with the Pad\'e of $\epsilon$ expansion in the region $3< D< 6$ is improved. The critical  dimension $D_c$, where the scale dimension of scalar $\Delta_\phi$ is vanishing, is used for the improvement of Pad\'e.  For the understanding of the intersection point of zero loci of $3\times 3$ minors, $2\times 2$ minors are
investigated in details, which are connected through Pl\"{u}ker relations.
 \end{titlepage}

\newpage
\section{Introduction}
\vskip 2mm
The conformal field theory was developed long time ago \cite{Ferrara1973}, and the modern numerical approach was initiated by \cite{Rattazzi2008}. The recent studies by this conformal bootstrap method reach to  some remarkable results,  for 3D Ising model \cite{ElShowk2012,ElShowk2014}, Yang-Lee edge singularities \cite{gliozzi2013,gliozzi2014} , $O(N)$ models \cite{Kos2015,Kos2016,Nakayama2014,Krishnan2015} and self-avoiding walk  \cite{Shimada2016}. 

The briefly summary of  the determinant method for the conformal bootstrap theory is following.  The conformal bootstrap theory  is based on the conformal group $O(D,2)$, and the conformal block $G_{\Delta,L}$ is the eigenfunction of Casimir differential operator $ \tilde D_2$.
The eigenvalue of this  Casimir operator is $C_2$, 
\ba\label{Casimir}
&&\tilde D_2 G_{\Delta,L} = C_2 G_{\Delta,L},\nonumber\\
&&C_2 = \frac{1}{2}[\Delta (\Delta - D) + L( L + D - 2)].
\ea

The solutions of the Casimir equation have been studied  \cite{James1968,Koornwinder1978,Dolan2004}. The conformal block $G_{\Delta,L}(u,v)$ has two variables $u$ and $v$, which denote the cross ratios, $u= (x_{12}x_{34}/x_{13}x_{24})^2$ , $v=(x_{14}x_{23}/x_{13}x_{24})^2$, where $x_{ij}= x_i-x_j$ ($x_i$ is two dimensional coordinate). They  are expressed as $u= z\bar z$ and $v=(1-z)(1-\bar z)$.
For the particular point $z=\bar z$,  the conformal block $G_{\Delta,L}(u,v)$ for spin zero (L=0) case  has a simple expression,
\be\label{hyper}
G_{\Delta,0}(u,v)|_{z=\bar z} = {(\frac{z^2}{1-z})^{\Delta/2}}  { }_3F_2[\frac{\Delta}{2},\frac{\Delta}{2},\frac{\Delta}{2}-\frac{D}{2}+1;\frac{D+1}{2},\Delta-\frac{D}{2}+1; \frac{z^2}{4(z-1)}]
\ee
The conformal bootstrap determines $\Delta$ by  the condition of the crossing symmetry $x_1\leftrightarrow x_3$.
For the practical calculations, the point $z=\bar z = 1/2$ is chosen  and by the recursion relation derived from Casimir equation of (\ref{Casimir}),
 $G_{\Delta,L}(u,v)$ at  $z=\bar z=1/2$ can be obtained \cite{ElShowk2012}.

The conformal bootstrap  analysis by a small size of matrix  has been investigated for Yang-Lee edge singularity with   accurate results 
of the scale dimensions by Gliozzi \cite{gliozzi2013,gliozzi2014}. For other models, this bootstrap method of determinant has been considered \cite{Fitzpatrick2016,Nakayama2016,Rychkov2015,Li2017}.  In this paper, we emphasize the importance of  the structure of minors along the Pl\"{u}ker relations as shown in appendix, and we apply it to Yang-Lee model  for the scale dimension of $\Delta_\phi$. The intersection point of the zero loci of $3\times 3$ minors was evaluated and it gave remarkable value of the critical exponent in \cite{gliozzi2013}. In this $3\times 3$ minors, if the value of the space dimension $D$ is given, the value of $\Delta_\epsilon=\Delta_\phi$ is determined from the
intersection point. However, if one goes to larger minors, for instance $4\times4$ or $5\times 5$ minors, one need the values of additional scale dimensions of operator product expansion (OPE).
In the study of such large minors, however, the values of $\Delta_\phi$ deviates from the $\epsilon$ expansion \cite{gliozzi2014}. This discrepancy should be improved by the determinant method or by the analysis of $\epsilon$ expansion, since the conformal bootstrap method should be consistent with $\epsilon$ expansion \cite{gliozzi2017,gliozzib2017}.  Recently, the consistent
results are obtained in O(N) vector model through Mellin amplitude \cite{Dey}.

We consider this discrepancy by repeating the evaluation of $3\times 3$ minors of Gliozzi \cite{gliozzi2013} and find that $3\times 3$ minors give accurate values which agree with the improved $\epsilon$ expansion (see Fig.8). The nature of the
determinant method is still not known, and  the convergence to the true value seems to be slow. As we mentioned before, we have to assume the values of scale dimensions of OPE, like spin 4, spin 6, etc. for the large minors. Unfortunately we don't know precisely these higher spin values at present. Therefore, we concentrate on  $3\times 3$ minors without any assumption of the other scale dimension of OPE.

The  four point correlation function is given by
\be
<\phi(x_1)\phi(x_2)\phi(x_3)\phi(x_4)> = \frac{g(u,v)}{|x_{12}|^{2\Delta_\phi}|x_{34}|^{2\Delta_\phi}}
\ee
and the amplitude  $g(u,v)$ is expanded as the sum of conformal blocks,
\be
g(u,v) = 1 + \sum_{\Delta,L} p_{\Delta,L} G_{\Delta,L}(u,v)
\ee
The crossing symmetry of $x_1\leftrightarrow x_3$ implies 
\be\label{crossing}
\sum_{\Delta,L} p_{\Delta,L} \frac{v^{\Delta_\phi}G_{\Delta,L}(u,v)- u^{\Delta_\phi}G_{\Delta,L}(v,u)}{u^{\Delta_\phi}-v^{\Delta_\phi}} = 1.
\ee

Minor method is consist of the derivatives at the symmetric point $z=\bar z= 1/2$ of (\ref{crossing}). 
By the change of variables $z=(a+ \sqrt{b})/2$, $\bar z= (a-\sqrt{b})/2$, derivatives  are taken about $a$ and $b$. 
Since the numbers of equations  become larger than the numbers of the truncated variables $\Delta$,
we need to consider the minors for the determination of the values of $\Delta$. The matrix elements of minors are expressed by,
\be\label{bootstrap}
f_{\Delta,L}^{(m,n)}= (\partial_a^m \partial_b^n \frac{v^{\Delta_\phi}G_{\Delta,L}(u,v)- u^{\Delta_\phi}G_{\Delta,L}(v,u)}{u^{\Delta_\phi}-v^{\Delta_\phi}})|_{a=1,b=0}
\ee
and the minors of $2\times 2$, $3\times 3$ for instance, $d_{ij}$, $d_{ijk}$ are  the determinants such as
\be\label{dijk}
d_{ij}= {\rm det}(f_{\Delta,L}^{(m,n)}),\hskip 3mm
d_{ijk} = {\rm det} ( f_{\Delta,L}^{(m,n)} )
\ee
where $i,j,k$ are numbers chosen differently from  (1,...,6), following the dictionary correspondence to $(m,n)$ as  $1\to (2,0)$, $2\to (4,0)$, $3\to (0,1)$, $4\to (0,2)$, $5\to (2,1)$ and $6\to (6,0)$.

In the Yang-Lee case, a fusion rule is  simply written as \cite{gliozzi2013}.
\be
[ \Delta_\phi ] \times [ \Delta_\phi ]  = 1 + [ \Delta_\phi ] + [ D,2 ] +  [\Delta_4,4 ] + \cdots
\ee

The scalar ($L=0$) term in the bootstrap equation of (\ref{bootstrap}) is denoted as
\be
vs0= \frac{v^{\Delta_\phi}G_{\Delta,0}(u,v) - u^{\Delta_\phi}G_{\Delta,0}(v,u)}{u^{\Delta_\phi}-v^{\Delta_\phi}}
\ee
We use the notation $(vsk)$ ($k$ is a number, related to the degree of the differential, and "v" is vector, "s" is a scalar) for the derivatives of $(vs0)$. rule in (\ref{bootstrap}).  For  $L=2$ we use ($vtk$), and for $L=4$ , we use $(vqk)$  notations. $(vxk)$ is for $L=6$.

The scalar ($L=0$) case is expressed by the hypergeometric function ($\Delta = \Delta_\phi$ for Yang-Lee edge singularity) 
\be\label{vs0}
vs0 = - 2^{-\Delta/2} ys + \frac{3 \times2^{-1-\Delta/2}}{2} ys + \frac{2^{-\Delta/2}}{2 \Delta} ys'
\ee
where
\be
ys = {}_3 F_2( \frac{\Delta}{2},\frac{\Delta}{2},\frac{\Delta}{2}-\frac{D}{2} + 1;\frac{\Delta}{2}+\frac{1}{2}, \Delta - \frac{D}{2}+1:  \frac{a^2}{8(-2+a)})
\ee
and $ys'$ is derivative of $ys$. We consider the derivative at a point $a=1$. 
\vskip 3mm
For instance, we have $2\times 2$ minors,
\be
d_{13} = {\rm det}(\begin{array}{cc}
vs1 & vs3\\
vt1 & vt3
\end{array} ),\hskip 3mm
d_{23} = {\rm det}(\begin{array}{cc}
vs2 & vs3\\
vt2 & vt3
\end{array} )\ee
 In this Yang-Lee edge singularity, we find the  $2\times 2$ minor, $d_{13}$, becomes zero at $D=6$ , and it provides exact values of $\Delta_\phi=2$. 
 
\section{$2\times 2$ minors}
\vskip 3mm
The intersection points of zero loci of $3\times 3$  minors are decomposed to $2\times 2$ minors at the critical dimensions $D=6$ for the free theory. In appendix (relation 1), this decomposition is shown.
 Among several $2\times 2$ minors, $d_{12}, d_{13}, d_{23},...$,  the most fundamental  minors may be  $d_{13}$ and $d_{23}$, which are made of lower derivatives of $a$ and $b$. Note that in Yang-Lee model, the constraint $\Delta_\epsilon =\Delta_\phi$ is taken, $2\times 2$ minor analysis corresponds to $3\times 3$ minor analysis of other model
 such as Ising model, in which $\Delta_\epsilon \ne \Delta_\phi$. From the point of the  truncation error in OPE as discussed in \cite{Li2017}, it is interesting to start from $2\times 2$ minors. Yang-Lee singularity is connected to the supersymmetry through the dimensional reduction of a branched polymer, where we have an exact relation of $\Delta_\epsilon = \Delta_\phi + 1$ \cite{Bashkirov2013,Hikami2017,Hikami2018}. We will discuss the dimensional reduction $D+2\to D$ behavior in Yang-Lee model in $2\times 2$ minors.

The $2\times 2$ minors are consist of the  two parameters, dimension $D$ and the scaling dimension $\Delta_\phi$, since  we have $\Delta_{\phi^2} = \Delta_\phi$ ($\Delta_\epsilon = \Delta_{\phi^2}$) for Yang-Lee case, due to the $\phi^3$ theory. 
\vskip 3mm
{\bf (i) Free field theory at D=6}
\vskip 2mm

The zero loci of $d_{13}$  gives exactly the scale dimension $\Delta_\phi = 2.0$ at $D=6$.
The minors for $vq$, $vx$ and $vs'$ ( with a scalar scale dimension  $\Delta'$ ) can be zero,
\be
{\rm det}(\begin{array}{cc}
vs1 & vq1\\
vs3 & vq3
\end{array} ) = 0,\hskip 3mm
 {\rm det}(\begin{array}{cc}
vs1 & vx1\\
vs3 & vx3
\end{array} )=0,\hskip 3mm
 {\rm det}(\begin{array}{cc}
vs1 & vs1'\\
vs3 & vs3'
\end{array} )=0\ee

For the value $\Delta_\phi=2$, they yields $Q(=\Delta_4)= 8$ and $\Delta_6= 10$, and $\Delta'=4$, respectively. These $\Delta_\phi,\Delta_4,\Delta_6,\Delta'$ are determined
exactly by $2\times 2$ minors of $d_{13}$. This is due to the free theory, i.e. no interactions between  the scale dimensions. It is known that the six dimension has algebraic identities, which are valid for a free theory in any dimension$D$ \cite{gliozzi2016}.

For other minor, for instance $d_{23}$, yields different value $\Delta_\phi = 2.158$ at $D=6$. These two different values of $\Delta_\phi$ are understood when we evaluate $3\times 3$ minors as shown in fig.2.
There are two fixed points of zero loci of  $3\times 3$ minors in the six dimensions.  The $3\times 3$ minors are denoted by
$d_{ijk}$ in (\ref{dijk}), for instance $d_{123}$ means
\be
d_{123}= {\rm det}(\begin{array}{ccc}
vs1 & vt1 & vq1\\
vs2 & vt2 & vq2\\
vs3 & vt3 & vq3
\end{array} ) 
\ee

\begin{figure}
\centerline{\includegraphics[width=0.6\textwidth]{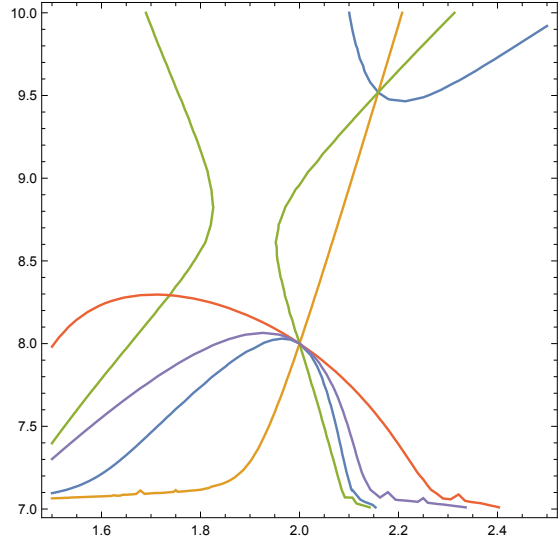}}
\caption{The intersection of zero loci of $3\times 3$  minors ($d_{123},d_{135},d_{134},d_{234},d_{235}$) for dimension $D=6$. The axis is (x,y) = $(\Delta_\phi, Q)$. The intersection point of five lines is $D=6, \Delta_\phi = 2$ with $Q=8$, which is decomposed to  $2\times 2$ minor $d_{13}$. The upper fixed point is related to the  minor $d_{23}$.}
\end{figure}
\vskip 2mm
{\bf (ii) D=3}
\vskip2mm
The zero loci of $2\times 2$ minor does not provide   good results except $D=6$ dimensions. This is due to the situation where the free theory  breaks down in general dimensions.  We have to consider the loci of $2\times 2$ minors of a finite small value due to the interactions, instead of zero value, for the non-trivial fixed point. The small non-vanishing value of the minor might have  important meaning for the bound of $\Delta_\phi$.
The loci of   two important $2\times 2$ minors $d_{13}$ and $d_{23}$ are shown in fig.2. in which the value of $d_{13}$ is changed by small values from -0.007 to 0.004, and the zero loci of $d_{23}$ is also shown in fig.3 as a guide line. The correct value of  $\Delta_\phi$ is realized in some finite value contours of $d_{13}$ for $D < 6$. Near D=6, the deviation from the zero loci of $d_{13}$  is expected since  $\Delta_\phi$ deviates from 2 as $\epsilon=6-D$ expansion shows; $ \Delta_\phi \simeq 2 -0.5555 \epsilon$.
\vskip 3mm
\begin{figure}
\centerline{\includegraphics[width=0.6\textwidth]{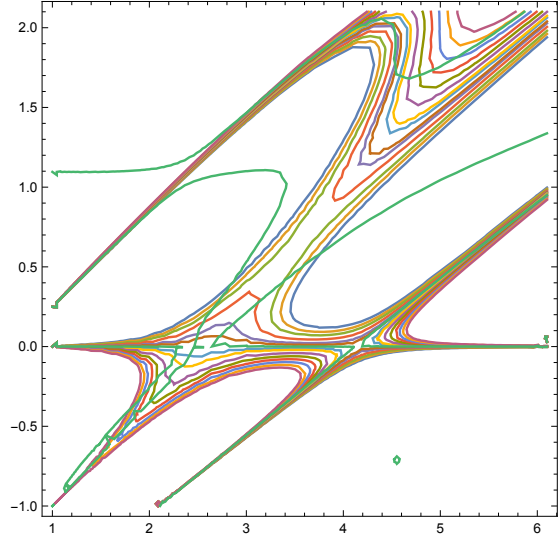}}
\caption{The  loci of $2\times 2$  minors $d_{13}$ with the values from -0.007 to 0.004 are shown, and $d_{23}=0$ is also shown by a green line.
The axis is  (x,y) = $(D, \Delta_\phi)$
.}
\end{figure}
In fig. 2, the curves of finite value of $d_{13}$ converge to a  curve starting from $\Delta_\phi= 0.25$ (at D=1) to 2.0 (at D=4). This curve resemble to the correct $\Delta_\phi$ contour if the space dimension is shifted
from $D$ to $D+2$. (The curve is translated to the right in fig.2 by two space dimensions by keeping the same value of $\Delta_\phi$).
This  dimensional reduction will be discussed in chapter 4.

The contour map of the finite value of $d_{13}$ shows  a valley shape, and some line is located at $D=2$ near the exact value $\Delta_\phi = - 0.4$  as shown in fig.3. In fig.3, the value of minor $d_{13}$ is 0.0034, which gives 
$\Delta_\phi=-0.4$. The cusp point moves when the value of minor $d_{13}$ is changed and it's trace may give the bound  of $\Delta_\phi$.

\begin{figure}
\centerline{\includegraphics[width=0.6\textwidth]{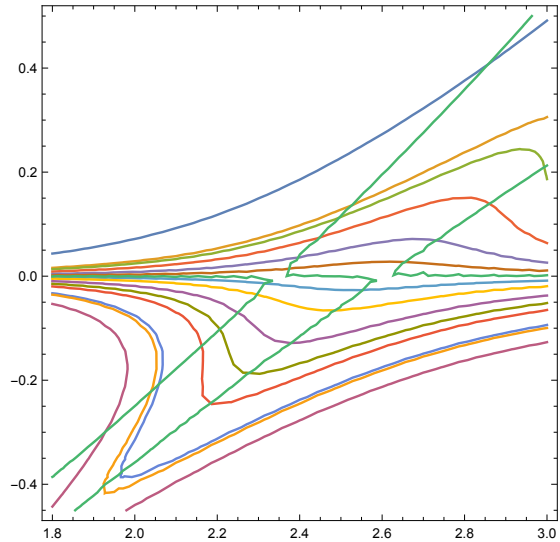}}
\caption{The closely looking contour map of fig.2 near the $\Delta_\phi=0$  loci of  minor $d_{13}$. The green line is the zero loci of $d_{23}$ as a  guide line. The axis is (x,y) = $(D, \Delta_\phi)$
.}
\end{figure}

\vskip 2mm
In fig.2 and fig.3, the contour maps of $d_{13}$ with small non-vanishing values, are shown.

\newpage

\section{Critical dimension for $\Delta_\phi = 0$}
\vskip 2mm
In the Yang-Lee edge singularity, the scale dimension $\Delta_\phi$ of $\phi$,  becomes zero at some dimension between 2 and 3, since it is$ -0.4$ in the two dimensions and approximately  0.2 in the three dimensions.
We call this dimension as critical dimension $D_c$, on which the scale dimension $\Delta_\phi$ vanishes. In the determinant method,  it is known that even small minors give accurate values of the scale dimensions \cite{gliozzi2013,gliozzi2014}.
We discuss here small determinants of $2\times 2$, $3\times3$ and $4\times 4$ matrices, for this critical dimension  $D_c$. 
\vskip 2mm

The figure of fig.2 for the zero or very small value of loci of $d_{13}$ shows $Z$ shape curves. This figure can be understood when we plot $\Delta_\epsilon$ v.s. $\Delta_\phi$ for various values of fixed dimensions. The minor of $d_{13}$ was considered as a function of $D$ and $\Delta_\phi$ up to now with the condition of $\Delta_\epsilon= \Delta_\phi$ due to Yang-Lee model. But in general we are able to consider $\Delta_\epsilon$ and $\Delta_\phi$ as two free parameters for $d_{13}$, and for yang-Lee case, we put $\Delta_\epsilon=\Delta_\phi$.

In fig.4, we plot the zero loci of $d[\Delta_\phi,\Delta_\epsilon]_{13}$ for different values of $D$. This minor then becomes usual one including the Ising case. The complicated contour of zero loci of $d_{13}$ is obtained specially around $D=4.4$, where it has three solutions for $\Delta_\epsilon = \Delta_\phi$. This solution corresponds to $Z$ shape around $(D,\Delta)= (4,1.5)$ in fig.2.

Also we find the line of $\Delta_\epsilon =\Delta_\phi$ of Yang-Lee condition goes through very near the zero loci of $d_{13}$, but it does not intersect. This corresponds to fig.2 with the loci of finite value of the $d_{13}$ around $D=3$.

 With some value of $D$, the contour in fig.4 (left down) and in fig. 5 intersect at the point $(\Delta_\epsilon,\Delta_\phi)=
(0,0)$. We find this critical dimension from fig.5 is
\be
D_c = 2.6199
\ee

\begin{figure}
\centerline{\includegraphics[width=0.6\textwidth]{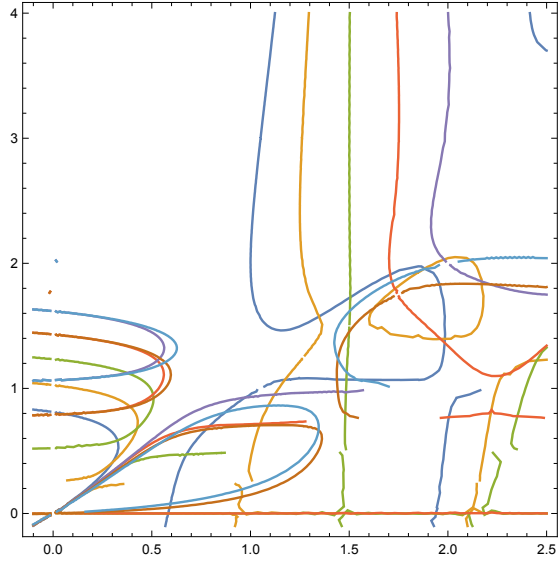}}
\caption{The zero loci of $d[\Delta_\phi,\Delta_\epsilon]_{13}$ for different dimensions.   The dark blue (D=6), red (D=5.5), green (D=5), yellow (D=4.5) and blue (D=4). The axis is (x,y) = $(\Delta_\phi, \Delta_\epsilon)$
.}
\end{figure}

\begin{figure}
\centerline{\includegraphics[width=0.6\textwidth]{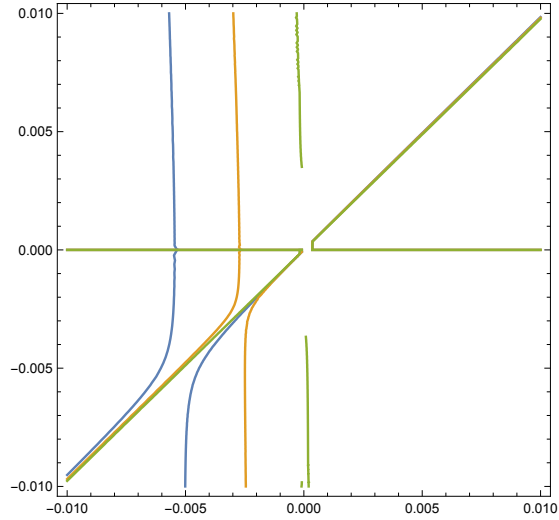}}
\caption{The contour plot of the zero loci of $d[\Delta_\phi,\Delta_\epsilon]_{13}$ for different dimensions, $D=2.6(blue), D=2.61(yellow),D=2.6199(green)$.  The axis is (x,y) = $(\Delta_\phi, \Delta_\epsilon)$
.}
\end{figure}

As seen in fig.3, for the small finite value of $d_{13}$,  the contour of $d_{13}$ is approaching the line $\Delta_\phi =0$, from both sides, above $\Delta_\phi >  0$ and $\Delta_\phi < 0$. The peak of contours are approaching to the critical dimension $D_c$. We consider the derivative of $d_{13}$ by the dimension $D$, and we estimate $D_c$ as 
the point of this derivative becomes zero. For $\Delta_\phi = 0.000001$, we obtain $D_c = 2.60574$. 
The derivative of $d_{13}$ by the space dimension $D$ gives the approximated solution of $D_c$ in the limit $\Delta_\phi \to 0$. The minor $d_{12}$ is proportional to $\Delta_\phi$, when $\Delta_\phi \to 0$ limit, since $vs1,vs3 \sim \Delta_\phi^2$, and $vt1,vt3 \sim \Delta_\phi^{-1}$.

We consider  how the minors become zero in the limit $\Delta_\phi \to 0$. For instance, at $D= 2.61$, we have for small $\Delta_\phi$,
\ba
&&vs1= 0.188 \Delta_\phi^2, \hskip 2mm vs2= 0.139\Delta_\phi^2,\hskip 2mm vs3= 0.225 \Delta_\phi^2\nonumber\\
&&vt1= \frac{1.055}{\Delta_\phi},\hskip 2mm vt2= \frac{0.667}{\Delta_\phi}, \hskip 2mm vt3= \frac{1.078}{\Delta_\phi}
\ea

The coefficients $p_\phi, p_t$ is obtained from
\be
(\begin{array}{cc}
vs0 & vt0\\
vs1 & vt1
\end{array}) (\begin{array}{c} p_\phi\\p_t \end{array} ) = (\begin{array} {c} 1\\ 0 \end{array})
\ee
Since $vs0=-\frac{1}{4}, vt0 = 0.7185/\Delta_\phi$ in the limit $\Delta_\phi \to 0$, we obtain  a solution as
$p_\phi=-4, p_t = a \Delta_\phi^2 {\rm log} \Delta_\phi$ ($a$ is some constant), which gives the value of central charge $C=0$.
With this choice of $p_t=p_{(D,2)}$, the central charge $ C$ becomes  for this critical dimension $D_c$,
$C = \Delta_\phi^2/p_t = 0.$
 This means that energy momentum tensor operator $T = \Delta_{(D,2)}$ can be neglected since OP coefficient of this operator $p_t$ becomes zero.
 This is well known $c$ catastrophe, which leads to logarithmic CFT \cite{Cardy2013}. When neglecting $\Delta_{(D,2)}$ term, and only considering $vs0$ in (\ref{vs0}), we obtain
 the critical dimension $D_c= 2.748632...$. However this value is too large compared to the expected value and might not be correct. We need more other operators to get  the
 correct value. We hope to get correct value of  the critical dimension $D_c$  by other sophisticated method by taking higher operators.

\vskip 3mm
{\bf {$[4\times 4$ minors]}}
\vskip 2mm
The numbers of zero loci of minors are ${}_6C_4= 6!/4!2!=15$,
We investigated the critical dimensions $D_c$, by changing $D$ between D=2.58 and D=2.59. Remarkably at $D=2.58953$, all 15 lines intersect at a single point in the contour map of $(Q, \Delta')$ under the condition $\Delta_\phi=0$. If we take it as the value of the critical dimension $D_c$, we find the following set of  results,
\be
D_c=2.58953, \hskip 2mm \Delta_\phi=0, \hskip 2mm Q= 4.40397,\hskip 2mm \Delta'=3.87127
\ee
\vskip 3mm
\begin{figure}
\centerline{\includegraphics[width=0.6\textwidth]{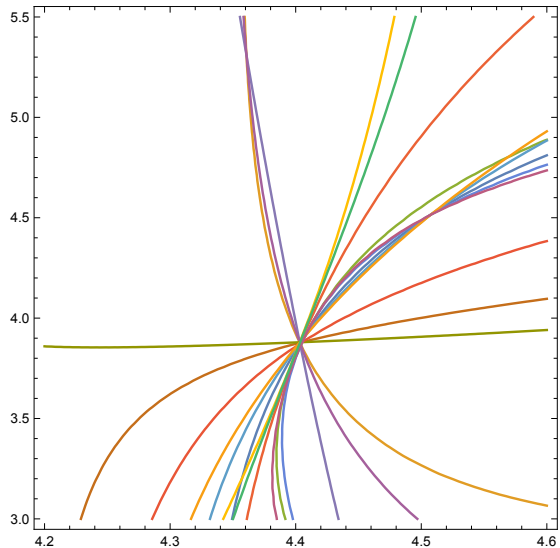}}
\caption{The intersection of zero loci of minors at  D=2.58953, $\Delta_\phi = 0$, and the axis is (x,y)=(Q,$\Delta'$)}
\end{figure} 
This critical dimension $D_c$ obtained $4\times 4$ minors is  close to $D_c=2.6199$  from $2\times 2 $ minor $d_{13}$ with the intersection point of $\Delta_\epsilon=\Delta_\phi=0$. In this figure 6,  all loci
coincide but if we  include more relevant operators, this value may change.
\vskip 2mm

For $\Delta_\phi=0$ case, in any dimension $D$, $d_{ij}$  can become zero.  When $\Delta_\phi=0$, by the definition of the scale dimension $\Delta_\phi$, two point correlation function $G(r) = 1/r^{2 \Delta_\phi}$ becomes a constant in the long range limit $r\to \infty$.

In the Yang-Lee edge singularity, the exponent of the density $\sigma$ is related to $\Delta_\phi$ as
\be
\sigma = \frac{\Delta_\phi}{D - \Delta_\phi}.
\ee
From this relation, $\sigma$ is vanishing for $\Delta_\phi=0$. This means the density is constant at the transition point. Several interesting systems are known in which  the density 
is constant  but
phase transition occurs. One example is a localization problem under the random potential.
\vskip 2mm
\section{Dimensional reduction}
\vskip 2mm
The zero loci of $2\times 2$ minor $d_{13}$ shows an interesting characteristic linear behavior for $D < 4$, which is approximated as $\Delta_\phi = (3D -2)/5$.
In Fig.7 , the shift of the zero loci of $d_{13}$ for $D\to D+2$ is shown (translation of two dimensions to the right direction in Fig.7). The blue line of Fig.7 for the zero loci of $d_{13}$   almost coincides with the red line of Yang-Lee edge singularity analyzed by Pad\'e approximation. The red line represents the result  of $3\times 3$ minors and it can be approximated as
\be\label{linear}
\Delta_\phi =\frac{3 D - 8}{5}
\ee
This expression satisfies the exact values of D=1 ($\Delta_\phi=-1$), D=2 ($\Delta_\phi= -0.4$) and D=6 ($\Delta_\phi$= 2).

This dimensional shift seems accidental, and only appears in a lower truncation, but there is a dimensional reduction which has been proven  rigorously.

\begin{figure}
\centerline{\includegraphics[width=0.6\textwidth]{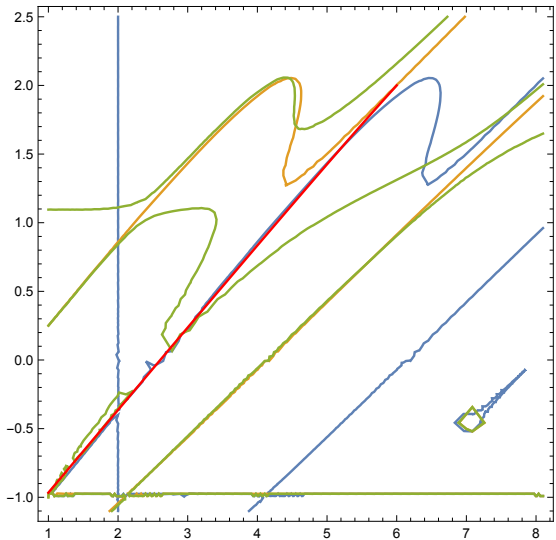}}
\caption{The $D\to D+2$ shifted line of the zero loci of $d_{13}$ (yellow line) is  shown by a line as blue color, which is consistent with the estimated Yang-Lee  line (red color) by Pad\'e analysis. The blue line and red line are almost same for $1 < D < 5.5$.  }
\end{figure}

 It is known that the branched  polymer in D+ 2 dimensions is equivalent to D dimensional Yang-Lee edge singularity for the critical phenomena. This dimensional reduction  has been explained by the supersymmetry \cite{Parisi1981}. Rigorous mathematical proof of this dimensional reduction is due to \cite{Imbrie2003}. We will study this dimensional reduction in a separate paper by the conformal bootstrap in a determinant method \cite{Hikami2017}. Yang-Lee edge singularity requires $\Delta_\epsilon (= \Delta_{\phi^2}) = \Delta_\phi$. 
 For a branched polymer due to dimensional reduction has a relation $\Delta_\epsilon = \Delta_\phi +1$. This relation is a manifestation of the supersymmetry of the system \cite{Bashkirov2013}.
 
Since the blue line  is very close to the red line in Fig.7, one can make estimation of the critical dimension $D_c$, in which $\Delta_\phi=0$.
The intersection of $d_{13}$ with $\Delta_\phi=0$ line  can be evaluated very precisely as  $D=0.5995471444$. Adding 2 to this value for the dimensional reduction, it gives the estimation of the critical dimension $D_c$ as $D_c = 2.5995471444$, which is very close to other estimations  of this paper. We have discuss this value in the previous section as $D_c = 2.6199$ as Eq.(18).
\vskip 2mm
\newpage

\section{$3\times 3$ minors and Pad\'e analysis}
\vskip 3mm
Up to now, we have discussed mainly of $2\times 2$ minors. If we take spin 4 operator, and its scale dimension $Q=\Delta_4$, we need the analysis for the intersection of zero
loci of $3\times 3$ minors $d_{ijk}$.

From the formula of minors in $3\times 6$ matrix in appendix,
we have Pl\"{u}ker formula such as (eq. (\ref{eq3}) in appendix)
\be
[126][345] -[123][456] + [124][356] - [125][346]= 0
\ee
For instance, at $D=3$, we find the zero loci of $d_{123}=[123]$, $d_{126}=[126]$, $d_{124}=[124]$ and $d_{346}=[346]$ intersect at a point, and above Pl\"{u}ker formula is
satisfied.

In table 1, we present the intersection of three zero loci of minors $d_{123}$, $d_{134}$ and $d_{124}$.
The value of $\Delta_\phi$ is obtained from the intersection point of the zero loci of minors. However, the intersection point is not only one, and we find at least three different
intersection points for each fixed dimension $D$.  In Table 1, we present with a parenthesis
such nearby different intersection point of three zero loci  of minors.
\vskip 3mm
{\bf Table 1}\\
\begin{tabular}[t]{|c|c|c|c|}
\hline
& $\Delta_\phi $& Q& $\Delta_\phi $ (Pad\'e)\\ 
\hline
D=3.0 & 0.174343 (0.187825) & 4.34106 (3.77124)  & 0.22995\\
D=3.5 & 0.499401 (0.500969)& 5.04195 (4.37556)& 0.53153 \\
D=4.0 & 0.823283 (0.815623)& 5.71152 (4.92283)& 0.83175 \\
D=4.5 & 1.13755 (1.12371)& 6.33395 (5.44645)  & 1.1300 \\
D=5.0 & 1.43807 (1.41987) & 6.91716 (5.95972) & 1.4255 \\
D=5.5 & 1.72469 (1.74682)& 7.46985 (6.46672)  & 1.7165 \\
D=6.0 & 2.0 & 8.0  & 2.0\\
\hline
\end{tabular}
\vskip 3mm



\vskip 2mm
In the $\epsilon = 6 - D$ expansion, $\Delta_\phi$ is known up to four loop \cite{Gracey2015}
\be
\Delta_\phi = 2 - 0.55555 x -0.0294925 x^2 + 0.021845 x^3 -0.0394773 x^4
\ee
where $x= \epsilon = 6 - D$.
Including the critical dimension $D_c$, for which $\Delta_\phi$ is vanishing, above expansion becomes
\be
\Delta_\phi = (6 - x - D_c)[ \frac{2}{6 - D_c} + (\frac{2}{(6 - D_c)^2} - \frac{0.55555}{6 - D_c}) x + O(x^2)]
\ee
Inserting the value of $D_c= 2.6199$ in section 4, it becomes
\ba
&&\Delta_\phi = (3.3801 - x)[0.59170 + 0.010695 x - 0.00556125 x^2 \nonumber\\
&&+ 0.00481755 x^3 - 0.0102541 x^4 + \cdots ]
\ea
This expansion is approximated by [2,2] Pad\'e method,
\be\label{Pade1}
\Delta_\phi = (3.3801 - x) [\frac{a_0 + a_1 x + a_2 x^2}{1 + b_1 x + b_2 x^2}]
\ee
with $a_0=0.59170, a_1= 2.3916, a_2= 1.0090, b_1=4.02385, b_2=1.6419$. 
The curve of this Pad\'e  is incorporated in Fig.6. 
\begin{figure}
\centerline{\includegraphics[width=0.6\textwidth]{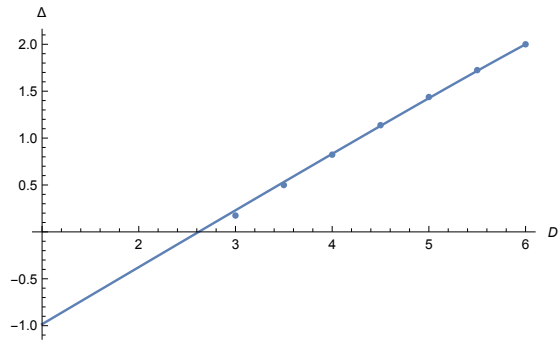}}
\caption{$\Delta_\phi$ is estimated by [2,2] Pad\'e with $D_c=2.6199$. The dots are the value of Table 1 for $3\times 3$ minor analysis of $\Delta_\phi$.}
\end{figure} 

\vskip 3mm
For the values of $\Delta_\phi$ of Yang-Lee singularity in D dimension, the previous results obtained by Gliozzi \cite{gliozzi2013} is very close to the values of Table 1, for 
instance, at D=4, $\Delta_\phi = 0.823$. In another evaluation with more primary operators \cite{gliozzi2014}, the values $\Delta_\phi=0.8466$ in D=4, $\Delta_\phi=1.455$ in D=5
are obtained, which are larger than the value of Table 1, and the comparison with $\epsilon$ expansion shows the apparent deviations as shown in Fig.5 of \cite{gliozzi2014}.
Our new analysis of Pad\'e with a fixed value at the critical dimension $\Delta_\phi=0$ seems more precise, and the  $\epsilon$ expansion and bootstrap determinant method
agree well in  each other as shown in Fig.8. One of aim of this paper is to clarify this point by the evaluation of the critical dimension $D_c$. Unfortunately, what we did
is the numerical estimate of $D_c$ by bootstrap determinant method and it may be approximation. It is important to obtain exact value of $D_c$, and the comparison with $\epsilon$ expansion becomes improved
by the exact value of $D_c$.

\newpage

\vskip 2mm
\section{Summary}
\vskip 2mm
We have considered Yang-Lee edge singularity in this article by bootstrap determinant method, initiated by Gliozzi. Although our results of the scale dimension $\Delta_\phi$ are same as $3\times 3$ determinant result of Gliozzi \cite{gliozzi2013}, we improved the consistency with the result of $\epsilon$ expansion. We find the basic $2\times 2$ minor $d_{13}$ is important for the determination of above scale dimensions through the analysis of the intersection points for the dimensions between one and six, although we used  the practical value of $\Delta_\phi$ is obtained by $3\times 3$
minors. We have shown the value obtained by the intersection of the zero loci of $3\times 3$ minors agrees with the result of $\epsilon$ expansion. The discrepancy between 
$\epsilon$ expansion and the determinant method  becomes small by the new  Pad\'e analysis with the introduction of the critical dimension $D_c$.

We obtained the critical dimension $D_c$ from the minor $d_{13}$ for $\Delta_\phi=0$ and $\Delta_\epsilon \to 0$ limits.
The intersection of the zero loci of $d[D,\Delta_\phi,\Delta_\epsilon]_{13}$ at the point $\Delta_\epsilon=\Delta_\phi=0$ gives the critical dimension 
$\Delta_c= 2.6199$.
 We have estimated critical dimension approximately by other methods, by the peak approaching of the finite $d_{13}$ $(D_c=2.6050)$,  the dimensional reduction value $(D_c=2.5995)$ and $4\times 4$ minor analysis $(D_c=2.589)$. 

We emphasize that the estimation of the critical dimension $D_c$ is practically useful  for the precise analysis of Yang-Lee edge singularity
between 2 and 6 dimensions by the Pad\'e   analysis of $\epsilon$ expansion \cite{Gracey2015}.  

We have discussed the dimensional reduction property in Yang-Lee model, on the fact that a branched polymer in D+2 dimension is equivalent to D dimensional Yang-Lee edge
singularity. 

\vskip 5mm
{\bf Acknowledgements}
\vskip 3mm
Author is thankful to Ferdinando Gliozzi for the discussions of the determinant method and  the useful suggestion of the critical dimension.  He also thanks Edouard Br\'ezin for discussion of the dimensional reduction problem. This work is supported by JSPS KAKENHI Grant-in-Aid 16K05491. The Mathematica11 of wolfram.com is acknowledged  for this research.
\newpage
{\bf \Large{Appendix (Pl\"{u}ker formula)} }
\vskip 3mm

\vskip 3mm

\vskip 3mm
{\bf{ Relation 1 ($3\times 3$ minors)}}
\vskip 2mm
For the determinant $d_{123}$, which is defined as
\ba
&&d_{123}= {\rm det} ( \begin{array}{ccc}
 vs1 & vs2 & vs3 \\
vt1 & vt2 & vt3 \\
vq1 & vq2 & vq3
\end{array} ) \nonumber\\
&&= (vq3 ){\rm det}( 
\begin{array}{cc}
 vs1 & vs2 \\
vt1 & vt2
\end{array} ) - (vq2) {\rm det}(\begin{array}{cc}
 vs1 & vs3 \\
vt1 & vt3
\end{array}) + (vq1) {\rm det}(  \begin{array}{cc}
 vs2 & vs3 \\
vt2 & vt3
\end{array} ), \nonumber\\
\ea
if the first determinant ($d_{12}$) and second minor $(d_{13})$ of r.h.s. are zero, then the third minor ($d_{23}$) should be  vanishing when $d_{123}=0$.
\vskip 2mm
\vskip 2mm
{\bf Relation 2; $2\times 4$ matrix}
\vskip 2mm
The $2\times 4$ matrix is denoted as
\be
M=(\begin{array}{cccc}
x_{11} & x_{12} & x_{13} & x_{14}\\
x_{21} & x_{22} & x_{23} & x_{24}
\end{array})
\ee
The Pl\"{u}ker relation is
\be\label{relation1}
[12][34] - [13][24] + [14][23] = 0
\ee
where the minor is denoted as $[ij] = (x_{1i}) (x_{2j}) - (x_{1j} )(x_{2i})$. The application of this formula to minors of our case can be taken as
\be
M=(\begin{array}{cccc}
vs1 & vs2 & vs3 & vs4\\
vt1 & vt2 & vt3 & vt4
\end{array})
\ee

Thus in our notation for $2\times 2$ minors,
\be
d_{12}= [12], \hskip 2mm d_{13}=[13],\hskip 2mm d_{14}=[14],\hskip 2mm
d_{23}=[23], \hskip 2mm d_{24}=[24],\hskip 2mm d_{34}=[34]
\ee
and we have an identity,
\be\label{Pluker2}
d_{12} d_{34} - d_{13} d_{24} + d_{14} d_{23} = 0
\ee
and it is represented by  tableau,
\be
\begin{array}{c} 
\fbox{1} \fbox{2}\\
\fbox{3} \fbox{4}
\end{array} 
-
\begin{array}{c} 
\fbox{1} \fbox{3}\\
\fbox{2} \fbox{4}
\end{array} 
+
\begin{array}{c} 
\fbox{1} \fbox{4}\\
\fbox{2} \fbox{3}
\end{array} 
= 0
\ee 
Note that the third  tableau is not  ordered in the increasing row and column (4 is larger than 3).  
Another application of this formula  is to take a following matrix,
\be\label{choice}
M=(\begin{array}{cccc}
vs1 & vt1 & vq1 & vx1\\
vs3 & vt3 & vq3 & vx3
\end{array})
\ee
In D=6 case,  we find $2\times 2$ minors made of above matrix are vanishing, $[12]=[13]= [14]=0$ for $\Delta_\phi=2,
\Delta_4= 8$ and $\Delta_6=10$. Therefore they satisfy the relation of (\ref{relation1}). In this $D=6$ case, we find  also for such values of $\Delta_\phi=2,\Delta_4=8,\Delta_6=10$, that 
other three sets of minors are  vanishing at $D=6$,
\ba
&&[34]= {\rm det}( \begin{array}{cc}
vq1 & vx1\\
vq3 & vx3
\end{array}) = 0, \hskip 3mm [24]= {\rm det}( \begin{array}{cc}
vt1 & vx1\\
vt3 & vx3
\end{array}) = 0, \nonumber\\
&&[23] ={\rm det}( \begin{array}{cc}
vt1 & vq1\\
vt3 & vq3
\end{array}) = 0
\ea
\vskip 3mm


{\bf Relation 3; $3\times 6$ matrix}\vskip2mm
(i) $3\times 6$ matrix  in bootstrap minor method is
\be
(\begin{array}{cccccc}
vs1 & vs2 & vs3 & vs4 & vs5 & vs6\\
vt1  &  vt2 & vt3 & vt4  & vt5  & vt6\\
vq1 & vq2 & vq3 & vq4 & vq5 & vq6
\end{array})
\ee

The Pl\"{u}ker relation is obtained by the mutual exchange of numbers;
\be\label{3pl}
[146][235]+ [124][356]-[134][256]+[126][345]-[136][245]+[123][456]=0
\ee
This identity is obtained from the first term to the second term by $(2\leftrightarrow 6)$, and from the second to third by $(2\leftrightarrow 3)$, etc. with the sign.
When we use following relations (obtained similarly),
\ba\label{eq3}
&&[126][345]-[123][456]+[124][356]-[125][346]=0\nonumber\\
&&[136][245]+[123][456]+[134][256]-[135][246]=0
\ea
From these equations, (\ref{3pl}) becomes
\be\label{tableau}
[146][235]= - 3 [123][456]-[125][346]+[135][246]
\ee
In our minor examples, we have
\be\label{15Pl}
d_{146} d_{235} = -3 d_{123} d_{456} - d_{125} d_{346} + d_{135} d_{246}
\ee

The terms of r.h.s. are ordered in the increasing row and column.
(ii) another application can be taken as

\be
(\begin{array}{cccccc}
vs1 & vt1 & vq1 & vx1 & vs1' & vs1''\\
vs2 & vt2 & vq2 & vx2  & vs2'  & vs2''\\
vs3 & vt3 & vq3 & vx3 & vs3' & vs3''
\end{array})
\ee
The Pl\"{u}ker relation is same as before.
In a tableau, (\ref{tableau})  is expressed as

\be
\begin{array}{c} 
\fbox{1} \fbox{4} \fbox{6}\\
\fbox{2} \fbox{3} \fbox{5}
\end{array} 
= - 3
\begin{array}{c} 
\fbox{1} \fbox{2} \fbox{3}\\
\fbox{4} \fbox{5} \fbox{6}
\end{array} 
-
\begin{array}{c} 
\fbox{1} \fbox{2} \fbox{5}\\
\fbox{3} \fbox{4} \fbox{6}
\end{array} 
+
\begin{array}{c}
\fbox{1} \fbox{3} \fbox{5}\\
\fbox{2} \fbox{4} \fbox{6}
\end{array}
\ee 
The left hand side is not ordered in  increasing column, but right hand side is a combination of ordered as both increasing row and column.

Pl\"{u}ker relation is expressed by two row tableau \cite{Bruns1988}, and for $m\times n$ matrix, the formula becomes
\be
\sum_{i_1< \cdots < i_t, i_{t+1}, \cdots < i_s} \sigma (i_1,...,i_s) [a_1,...,a_k,c_{i_1}, ..., c_{i_t}][c_{i_{t+1}},...,c_{i_s},b_l,....,b_m] = 0
\ee
where $(1,...,s)=(i_{1},...,i_s)$ and $a_j,b_{j'},c_{j''} \in (1,...,n)$ and $s=m-k+l -1 > m, t= m-k>0$.

 These tableau representations suggest the algebraic structures like the character expansion.

\newpage

\end{document}